\documentclass{article}          

\begin {document}

\title {APPROXIMATE GRAVITATIONAL FIELD OF A ROTATING DEFORMED MASS}

\author{J.L. Hern\'andez-Pastora\thanks{Departamento de Matem\'atica Aplicada.  E.T.S. Ingenier\'\i a Industrial de B\'ejar. Universidad de Salamanca.  Salamanca, Espa\~na.  }}

\date{} \maketitle

\begin{abstract}
A new approximate solution of vacuum and stationary Einstein field
equations is obtained. This solution is constructed by means of a
power series expansion of the Ernst potential in terms of two
independent and dimensionless  parameters representing the quadrupole
and the angular momentum respectively.  The main feature of the
solution is a suitable description of small deviations from spherical
symmetry through perturbations of the static configuration and the
massive multipole structure by using those parameters.  This quality
of the solution  might eventually provide relevant differences with
respect to the description provided by the Kerr solution.
\end{abstract}

\vskip 1cm PACS numbers: 04.25.Nx, 04.20.Cv, 04.20.-q

\newpage

\section{Introduction}

Within the context of axisymmetric Einstein vacuum field equations, 
the good qualities of the Kerr solution \cite{kerr} are well known and indeed
this solution is the essential reference in the stationary domain, as is the
Schwarzschild solution in the static one.  We are dealing with an
exact solution which is very simple and has a  very interesting multipole moment
structure in order to describe the gravitational field of compact
bodies. The Kerr solution is written in terms of  only one parameter,
namely $a$, which represents the angular momentum ($J$) per unit mass. 
$M$ being the mass - i.e. the order zero of the multipole moments-, all the
other moments  are  proportional to a power of this parameter $a$
equal to  each multipole order, \cite{mk}, \cite{fhp}
\begin{equation}
M_k = M (i a)^k
\label{kerr}
\end{equation}
It is worthwhile mentioning that all multipole moments of the Kerr
solution equal the corresponding coefficient appearing in the
expansion of its Ernst potential on the symmetry axis. This
characteristic makes it easier to calculate the multipole moments of the
solution by using the FHP algorithm \cite{fhp}.

All these reasons, among others, have led to this solution becoming the
one most used for a broad range of purposes concerning the physical
behaviour of test particles or properties of the gravitational fields
of celestial sources.

By means of the multipole moments of any solution one can obtain
relevant information about the physical properties of the
gravitational field, and, for instance, the multipole structure (\ref{kerr}) allows one
to estimate the relative importance of the successive multipole moments
in terms of the parameter $a$, mainly for the sub-extreme case ($J <<
M$).

Nevertheless, since the Kerr solution is defined with  only one arbitrary
parameter, it is not possible to distinguish the different contributions
from  the quadrupole and angular momentum  terms, respectively, to any
kind of physical event described by this space-time.  It would be
very useful to question how different a  solution is with respect to
the Schwarzschild solution in order to know which  different
contributions  (or the new physics involved)  are introduced by a
 new solution when this  has no spherical symmetry and this space-time is no longer
static.  To make such a description, a  solution with more than one
parameter, themselves themselves related to different multipole
moments, allows one to  handle them independently.

In this sense, some authors \cite{y1}, \cite{y2} have supplied static
and stationary solutions of the Einstein vacuum field equations, with
prescribed multipole moments, that attempt to describe slight deviations
from the space-time with spherical symmetry.  The Monopole-Quadrupole
solution $MQ$ \cite{y1} is an exact axisymmetric and static solution
constructed in power series of a parameter $q$ (dimensionless
quadrupole moment), whereas the $MJ$ solution \cite{y2} is a stationary
approximate solution describing  the space-time of a mass with angular
momentum and is obtained by an expansion on the symmetry axis of the
Ernst potential in power series of a dimensionless parameter ${\cal
J}$. This solution represents successive corrections to the spherical
symmetry due to the rotation, the sum of the series being the pure
Monopole-Dynamic dipole vacuum solution.

One way to provide the physical content to these solutions consists in
establishing a link between  their multipole moments and quantities
measured from well defined and physically reasonable experiments. This was
 the purpose of  paper \cite{gyros}, in which  we established such a
link by calculating the total precession per revolution of gyroscopes
circumventing the symmetry axis.

As a second result, the possibility emerges of comparing different axisymmetric
solutions in term of an observable quantity (angle precessed by the
gyroscope). In particular, differences in  the contribution
of the quadrupole moment and the angular momentum for the Erez-Rosen
solution \cite{er} and Kerr solution \cite{kerr} were obtained with
respect to the $MQ^{(1)}$ (see \cite{y1}, \cite{gyros} or a brief comment
below) and $MJ$ solutions respectively.

Some works \cite{colapso} have addressed  the
description of non-spherical collapse and the relevance of this
circumstance on the fate of the  collapse; and other works \cite{thermal} study the influence of thermal conduction on the evolution of a self-graviting system out of hydrostatic equilibrium. Most of the life of a star (at any stage of evolution), may be described on the basis of the quasi-static approximation (slowly evolving regime), and this is so, because most relevant processes in star interiors take place on time scales that are usually much longer than the hydrostatic time scale \cite{thermal1}, \cite{thermal2}. Instead of following the evolution of the system a long time after its departure from equilibrium, the system is evaluated inmediately after such a departure. Here ' inmediately' means on a time scale of the order of the thermal relaxation time, before the stablishment of the steady-state resistive flow. In doing so, it is avoided the introduction of numerical procedures which might lead to model-dependent conclusions, and for that evaluation of the system, an analytic solution becomes fundamental to make use of suitable approximations; in particular, it is very important to have a solution constructed in  a suitable way to describe the
contributions to these scenarios of both the rotation and the quadrupole
deformation separately.  On the other hand, however, it is only obtained indications about tendency of the object and not a complete description of its evolution. Then, it should be clear that, for sure, the numerical methods used to solve the complicated mathematical problem arising at the study of non-spherical colapse scenarios are much better to obtain a complete description.

As is well knwon, the Weyl metrics \cite{weyl2} are static axysymmetric solutions to vacuum Einstein equations which are given by the line element,
\begin{equation}
ds^2 = -e^{2 \Psi} dt^2+e^{-2 \Psi} (e^{2 \gamma}(d \rho^2+dz^2)+\rho^2 d \phi^2)
\end{equation}
where metric functions have to satisfy
\begin{equation}
\Psi_{,\rho \rho}+\rho^{-1} \Psi_{,\rho}+ \Psi_{,zz}=0
\label{laplace}
\end{equation}
and
\begin{equation}
\gamma_{,\rho} = \rho (\Psi_{,\rho}^2-\Psi_{,z}^2) \quad , \quad \gamma_{,z} = 2 \rho \Psi_{,\rho}\Psi_{,z}  .
\label{gamma}
\end{equation}
Observe that (\ref{laplace}) is just the Laplace equation for $\Psi$ (in the Euclidean space), and furthermore it represents the integrability condition for (\ref{gamma}), implying that for any''Newtonian'' potential we have a specific Weyl metric. Another interesting way of writting the general solution of Laplace equation, representing an asymptotically flat behaviour, was obtained by Erez-Rosen \cite{er} and Quevedo \cite{weyl}, integrating equations (\ref{laplace}), (\ref{gamma}) in prolate spheroidal coordinates. A sub-family of Weyl solutions has been obtained by Gutsunayev and Manko \cite{manko}, starting from the Schwarzschild solution as a seed solution.

In \cite{y1}, \cite{luisbarreto} the $MQ$ solution was presented
in the following way
\begin{equation}
\displaystyle{\Psi_{M-Q} = \Psi_{q^0}+ q \Psi_{q^1}+ q^2 \Psi_{q^2}+
\dots = \sum_{\alpha=0}^{\infty} q^{\alpha} \Psi_{q^{\alpha}}} \ ,
\end{equation}
where the zeroth order $\Psi_{q^0}$ corresponds to the Schwarzschild
solution,  and $\Psi_{q^i}$ are series that can be summed to obtain
exact solutions of the Gutsunayev-Manko family \cite{manko}  and
the Erez-Rosen family as well \cite{weyl}.   It appears that each power in
$q$ adds a quadrupole correction to the spherically symmetric
solution, the first one being referred to as $MQ^{(1)}$ \cite{y1}. It should be observed that due to the linearity of Laplace equation, these corrections give rise to a series of exact solutions, and so, the power series of $q$ may be cut at any order, and the partial summatory, up to that order, gives an exact solution representing a quadrupolar correction to the Schwarzschild solution.

The existence of so many different (physically distinguishable \cite{giros}) Weyl solutions gives rise to the question: which among Weyl solutions is better entitled to describe small deviations from spherical symmetry?. Although it should be obvious that such a question does not have a unique answer ( there is an infinite number of ways of being non-spherical, so to speak), we shall invoke a very simple criterion, emerging from Newtonian gravity, in order to choose that solution: for example, massive multipole moments of an ellipsoid of rotation, with homogeneous density, mass $M$ and axes $(a,a,b)$ read:
\begin{eqnarray}
M_{2n} & = & \displaystyle{\frac{(-2)^n 3 M a^{2n} \epsilon^n (1-\epsilon/2)^n}{(2n+1)(2n+3)}} \quad , \quad \epsilon \equiv (a-b)/a, \\
M_{2n+1} & = & 0
\end{eqnarray}
because of the factor $\epsilon^n$, this equation clearly exhibits the progressive decreasing of the relevance of multipole moments as $n$ increases.
Same conclusion can be obtained for the multipole moments of rotation associated to a  homogeneous distribution of charge, with total charge $Q$ and ellipsoidal geometry with axes $(a,a,b)$ \cite{tesis}:
\begin{eqnarray}
J_{2n} & = &  0\\
J_{2n+1} & = & \displaystyle{\frac{(-2)^n 3 Q a^{2n+2} \epsilon^n (1-\epsilon/2)^n \Omega}{(2n+1)(2n+3)(2n+5)}}
\end{eqnarray}
$\Omega$ being the angular velocity of the rigid rotating distribution.

Thus, in order to describe small departures from sphericity, by means of a solution of Einstein equations, it would be requiered a solution  whose multipole structure shares the property mentioned above. For that reason,
the  $MQ$ solution is particularly suitable for the study of
perturbations of spherical symmetry. The main argument to support this
statement is based on the fact that the previously known Weyl \cite{weyl}
metrics (e.g. Gutsunayev-Manko, Manko, $\gamma$-metric \cite{gymas},
etc) present a drawback when describing quasi-spherical space-times. It consists of the fact that its multipole structure is such that all the moments higher than
the quadrupole are of the same order as the quadrupole moment. Instead, as it is intuitively clear and as it is shown above for the classical case, the relevance of such multipole moments should decrease as we move from lower to higher moments, the quadrupole moment being the moste relevant for a small departure from sphericity.

In \cite{geod}, the behaviour of geodesics is compared with the
spherically symmetric situation, throwing light on the sensitivity of the
trajectories to deviations from spherical symmetry. The change of sign in the proper radial acceleration of
test particles moving radially along the  symmetry axis, close to the $r=2
M$ surface, and related to the quadrupole moment of the source  deserves  
particular
attention.

The ultimate aim of the above works \cite{y1}, \cite{y2} (see
\cite{tesis} for general expressions), and the main contribution of the solution presented here, is to obtain a hierarchy of
solutions describing pure multipole moment space-times. 
Since 1918,  when Schwarzschild published its  static solution of the
Einstein vacuum  field equations, one of the  more active topics of
relativistic scientists is devoted to obtain exact solutions to those
equations. So, up to now a lot of works have contributed not only with
new solutions but developping several techniques to construct new
solutions, specifically in the axial symmetry case, nevertheless with
arbitrary, in general, physical content.  Afterwards,  main aim
aroused  was to  get  the correct  physical interpretation of the
solutions  obtained. Some of these solutions are already known and
studied but in general there is no specific method which allows us
both to understand  which is the physical relevance of a new solution
as well as to construct solutions with prescribed physical behaviour
beyond of the symmetry properties of the problem.
In \cite{y1}, \cite{y2}, a  generalization of the classical gravitational field to 
General Relativity theory was planned,  by describing the solution defined as
the sum of different  pure multipole contributions, (following an analogy of the solutions
for the classical potential), constructed by means of suitable harmonic
solutions of the Laplace equation. 
A recent work,
\cite{mag2}, establishes the reciprocal relationship between static
solutions with axial symmetry  and any prescribed  multipole
structure. The authors in \cite{mag1}, \cite{mag2} show how to obtain 
the Weyl moments required for constructing the pure multipole
solutions sought. Although these works are highly relevant, we have not yet a generalization of those results to the stationary general case, and this is the motivation of the solution presented here. 

The following step to the $MQ$ solution is the pure multipole solution,
which has only mass, angular momentum and quadrupole moment,
referred to hereafter as the $MJQ$ solution. Here we present  an
approximation to this solution by means of an expansion of the
corresponding  Ernst potential. It will be shown that the resulting
solution is a very good candidate to describe corrections to the
spherical and static configuration due to the rotation and the
quadrupole deformation independently.

For this exterior metric that is presented here (MJQ-solution) it would be very interesting to construct a source. The motivation for this is twofold: on the one hand, it is always interesting to propose bounded and physically reasonable sources of gravitational fields,  which may serve as models of compact objects. On the other hand, spherical symmetry is a common assumption in the study of compact self-gravitating objects (white dwarfs, neutron stars, black holes). Therefore it is pertinent to ask, how do small deviations from this assumption , related to any kind of perturbation (e.g. fluctuations of the stellar matter, external perturbations, etc.), affect the properties of the system?. However, for sufficiently strong fields, in order to answer to this question it is necessary to deal with non-spherically symmetric solution of Einstein equations. Work  towards this way is in progress, and, for example, in \cite{luisbarreto} an interior solution for the $MQ^{(1)}$ solution was already obtained.

The paper is organized as follows: In the next section, all relevant
equations are given; we develop the construction  of the approximate  $MJQ$ 
solution and  analyze of the behaviour of the solution
for limiting cases. Reasons for  appreciating the physical relevance of
this solution and its meaningful contribution to the description of
small deviations from spherical symmetry are given.  Finally, a
discussion of the results is presented in the last section, as well as
some comments on forthcoming works and the proposals therein.

\section{The MJQ Solution}
\subsection{The approximation of the solution}

In 1968 Ernst \cite{ern} showed a simplification of the field equations derived from the vacuum stationary and axysymmetric Einstein equations. In order to do that, Ernst made use of a variational principle on the Lagrange function constructed with  the metric functions $f$ and $\omega$ of the line element describing those metrics, which in Weyl canonical coordinates \cite{weyl2} reads as follows:
\begin{equation}
ds^2 = -f (dt - \omega d \phi)^2+f^{-1}[e^{2 \gamma} (d \rho^2+d z^2)+\rho^2d \phi^2]
\end{equation}
being $f$, $\omega$ and $\gamma$ functions only depending on $(\rho, z)$.

A very simple way to obtain the Ernst equations, is by means of a suitable rewrite of the field equations; a redefinition of the metric function $\omega$ is made by introducing a scalar function $W$ whose existence is guaranteed by one of the Einstein field equations\footnote{It can be proved that the gradiant of this scalar $W$ is exactly the projection of the twist of the time Killing vector, describing  stationarity, over the 3-dimensional quotient manifold}.
Then, the Ernst potential $E$ is defined as a complex function whose real and imaginary part are the metric functions $f$ and $W$ respectively, the real part representing the norm of the Killing vector.

On symmetry axis, the potential $\xi$, which is the transformation $\xi
{\displaystyle \equiv \frac{1-E}{1+E}}$ from the original Ernst
potential, can be expanded by means of a power series of the inverse Weyl's canonical coordinate $z$ as follows:
\begin{equation}
\xi (\rho=0,z)= \sum_{n=0}^{\infty} m_n z^{-(n+1)} \,
\label{xiaxis}
\end{equation}
where $\rho$ represents the Weyl radial coordinate.

In \cite{y2} a method was proposed for obtaining an approximate
solution of the stationary vacuum Einstein field equations ($MJ$-solution  or pure
monopole-dynamic dipole solution) by means of an expansion of the
Ernst potential $\xi$ in power series of a dimensionless parameter, ${\cal
J}$, directly related to the angular momentum of the solution. The
expansion to describe that solution ($MJ$) is quite appropriate, since
it is known that its Ernst potential  on the
symmetry axis can be written in terms of successive powers of that
parameter. This conclusion is obtained from the fact that
 Fodor, Hoenselaers and Perj\'es \cite{fhp} have developed an algorithm (FHP) which allows us to calculate the Geroch \cite{geroch} and Hansen \cite{mk} relativistic multipole moments, related to a vacuum stationary axisymmetric solution, in terms of the  coefficients ($m_n$) of the expansion on the symmetry
axis of the Ernst potential $\xi$ (\ref{xiaxis}). Both the result obtained up to multipole order 10 by these authors, as well as the calculations we have carried out up to order 20 \cite{y2}, show that the relation between multipole moments and coefficients $m_n$ is triangular: that is to say, the multipole moment and the the corresponding coefficient $m_n$ at every order differ in a certain combination of lesser order $m_k$ coefficients. Therefore, these relations can be suitably inverted, and it  enable us to determine unequivocally the coefficients $m_n$ which define the Ernst potential $\xi$ in terms of the multipole moments for any given solution. For example in \cite{y2} was obtained the coefficients which characterize the Ernst potential of the solution having only massive monopole (the mass) and dynamic dipole (angular momentum).

We are now looking  for the stationary  axisymmetric space-time whose
multipole moment structure only has angular momentum and quadrupole
moment in addition to the mass. By imposing this condition on the
coefficients of the series expansion of the Ernst potential along the
symmetry axis (\ref{xiaxis}) we obtain the following expressions (only the fists ones are shown, but they have been calculated up to order 20):
\begin{eqnarray}
\nonumber m_0 &=& M \\ \nonumber
m_1 &=& i J \\ \nonumber
m_2 &=& Q \\ \nonumber
m_3 &=& 0 \\ \nonumber
m_4 &=& \frac 17 M J^2+ \frac 17 M^2 Q \\ \nonumber
m_5 &=& - \frac {i}{21} J^3- i \frac{8}{21} MJQ \\ 
m_6 &=& \frac {1}{21} M^4 Q+ \frac {1}{21} M^3 J^2- \frac{51}{231} M Q^2 - \frac{23}{231} Q J^2 \\ \nonumber 
\label{emesfhp}
\end{eqnarray}
$J$, $Q$ and $M$  being the
angular momentum, the quadrupole moment and the mass of the solution
respectively. This result leads to the Ernst potential $\xi$ on the symmetry axis of the following type
\begin{equation}
\xi(\rho=0,z) = \sum_{n=0}^{\infty} {\cal J}^n \sum_{k=0}^{\infty} q^k
F(n,k) \ ,
\label{estructura}
\end{equation}
where ${\cal J} \displaystyle{ \equiv i \frac {J}{M^2}}$ and $q
\displaystyle{\equiv \frac{Q}{M^3}}$, and the  double index function $F(n,k)$ is a series that
contains information about the dependence on the Weyl coordinate, $z$,
along the axis with the corresponding numerical factor derived from
the coefficients $m_n$ (\ref{emesfhp}).  For instance, it can be seen that $F(0,0) =
\displaystyle{\frac Mz}$ and $F(1,0) = \displaystyle{\frac {M^2}{z^2}}$.
Furthermore,  by considering only $k=0$ in the double sum of (\ref{estructura}), then the expression for the $MJ$ solution can be obtained.

Since we are interested in  the description  of slight deviations from
spherical symmetry, we shall take into account the leading terms in
the expansion (\ref{estructura}) assuming that  $q$, ${\cal J} << 1$,
and neglecting terms involving powers or products of both parameters.

Thus, with these considerations the approximate solution up to
order $q$ and ${\cal J}$, describing the space-time with mass, angular
momentum and quadrupole  moment, is given by the following Ernst
potential on the symmetry axis:
\begin{equation}
\xi(\rho=0,z) = \xi_0(\rho=0,z)+\xi_1(\rho=0,z) {\cal
J}+\xi_2(\rho=0,z) q \ ,
\label{xijq}
\end{equation}
where
\begin{eqnarray}
&\xi_0(\rho=0,z)&=\displaystyle{\frac Mz}\\
&\xi_1(\rho=0,z)&=\displaystyle{\frac{M^2}{z^2}}  \label{orden1}\\
F(0,1) \equiv &\xi_2(\rho=0,z)& = \sum_{j=1}^{\infty} m_{2j}
z^{-(2j+1)} = \sum_{j=1}^{\infty} \hat\lambda^{2j+1} F_{2j}^q \ ,
\label{orden2eje}
\end{eqnarray}
with $\hat\lambda \displaystyle{\equiv \frac Mz}$ and $F_{2j}^q
=\displaystyle{ \frac {15}{(2j+3)(2j+1)(2j-1)}}$ because we know from
the FHP algorithm  (see coefficients $m_n$ of (\ref{emesfhp})) that the 
$MJQ$ solution, up to order $q$ and ${\cal J}$,  is defined by the following coefficients:
\begin{equation}
m_{2j+1}=0 \quad , \quad m_{2j}=F_{2j}^q M^{2j+1} q \quad , \quad
\forall j \geq 1 \ .
\end{equation}

Now, in what follows, and according to the scheme developed in \cite{y2}, we proceed  to
solve the Ernst equation \cite{ern},
\begin{equation}
(\xi \xi^{\star}-1) \triangle \xi = 2 \xi^{\star} (\nabla \xi)^2
\label{ernst}
\end{equation}
to each order in ${\cal J}$ and $q$
for the full Ernst potential $\xi$, with the following form
\begin{equation}
\xi = \xi_0 + \xi_1 {\cal J} + \xi_2 q
\end{equation}
The set of constants appearing in the general solution of
(\ref{ernst}) to each order, will be determined by restricting the solution
on the symmetry axis and comparison with  expression (\ref{xijq}).
The zero order in ${\cal J}$ and $q$ gives
\begin{equation}
(\xi_0^2-1) \triangle \xi_0 = 2 \xi_0 (\nabla \xi_0)^2 \ ,
\label{ernstzero}
\end{equation}
and we take $\xi_0$ to be the Ernst potential corresponding to the
Schwarzschild solution.

The first order in parameter ${\cal J}$, i.e., $\xi_1$ is already
obtained in \cite{y2} by solving the equation
\begin{equation}
(\xi_0^2-1) \triangle \xi_1-4 \xi_0 \nabla\xi_0 \nabla\xi_1+2 \xi_1
(\nabla\xi_0)^2=0 \ ,
\end{equation}
and by imposing as a neighbourhood condition the known behaviour of the
solution on the symmetry axis (\ref{orden1}). It should be noted
that the approximate solution to this order, which in prolate
coordinates is written 
\begin{equation}
\xi = \xi_0 + \xi_1 {\cal J} = \frac 1x + \frac{y}{x^2} {\cal J} \ ,
\label{kerr1orden}
\end{equation}
is the same as that arising from the expansion of the Kerr solution on
parameter ${\cal J}= i \displaystyle{\frac aM}$ up to first order.

The first order in parameter $q$ ($\xi_2$) must fullfils the following
equation:
\begin{equation}
(\xi_0^2-1) \triangle \xi_2 - 4 \xi_0\nabla \xi_0\nabla\xi_2 +2 \xi_2
(\nabla \xi_0)^2 \frac{\xi_0^2+1}{\xi_0^2-1}= 0 \ .
\end{equation}
This homogeneous equation can be simplified by means of the following
redefinition of the function $\xi_2$:
\begin{equation}
\zeta_2 \equiv \frac{\xi_2}{\xi_0^2-1} \ ,
\end{equation}
which leads to the Laplace equation for the function $\zeta_2$. It can
be easily checked that, as is known, the equation $\triangle \zeta_2=0$
is separable and the general solution  with regular behaviour on the
symmetry axis ($y=\pm1$) as well as in the neighbourhood of infinity, affords:
\begin{equation}
\zeta_2 = \sum_{n=0}^{\infty} h_n^q Q_n(x) P_n(y) \ ,
\end{equation} 
and therefore:
\begin{equation}
\xi_2 =\frac{1-x^2}{x^2} \sum_{n=0}^{\infty} h_n^q Q_n(x) P_n(y) \ ,
\label{orden2}
\end{equation} 
where $\{x,y\}$ are prolate coordinates; $h_n^q$ are constants of
integration, and $Q_n(x)$, $P_n(y)$ are associated Legendre functions
of the second kind and Legendre polynomials respectively.

In order to compare the restriction on the symmetry axis of this
result (\ref{orden2}) with expression (\ref{orden2eje}), we rewrite
(\ref{orden2eje}) as follows:
\begin{equation}
\xi_2(\rho=0,z) = \frac{M^2-z^2}{z^2}\big[\frac{1}{\hat\lambda^2-1}
\xi_2(\rho=0,z)\big] \ ,
\end{equation}
and by carrying out an expansion on the parameter $\hat\lambda$ we have:
\begin{equation}
\xi_2(\rho=0,z) = \frac{z^2-M^2}{z^2}
\big[\sum_{i=0}^{\infty}\hat\lambda^{2i}\sum_{j=1}^{\infty}\hat\lambda^{2j+1}
F_{2j}^q] = \frac{z^2-M}{z^2} [\sum_{j=1}^{\infty}\hat\lambda^{2j+1}
I_j^q\big] \ ,
\end{equation}
with the definition $I_j^q \equiv \displaystyle{\sum_{i=1}^j
F_{2i}^q}$.  By using Lemma 4 of the Appendix in \cite{y2} we rewrite this
expression in terms of associated Legendre functions of the second kind as
follows:
\begin{equation}
\xi_2(\rho=0,z) = \frac{z^2-M^2}{z^2} \sum_{j=1}^{\infty} I_j^q
\sum_{i=j}^{\infty} (4i+1) L_{2i,2j} Q_{2i}(1/\hat\lambda) \ ,
\end{equation}
$L_{2i,2j}$ being the coefficient of the Legendre polynomial of order $2i$,
with the  power $2j$ of the variable. Reordering the sums, we obtain:
\begin{equation}
\xi_2(\rho=0,z) = \frac{z^2-M^2}{z^2} \sum_{i=1}^{\infty}
Q_{2i}(1/\hat\lambda) (4i+1) \sum_{j=1}^{i} I_j^q L_{2i,2j} \ .
\label{ques}
\end{equation}

We now calculate $I_j^q$ for our case:
\begin{equation}
I_j^q = \frac{5j(j^2+2)}{(2j+3)(2j+1)} = \frac 54 -\frac {15}{8}
\frac{1}{2j+1}+\frac {15}{8} \frac{1}{2j+3}
\label{iq}
\end{equation}
and by comparing expression (\ref{orden2}) on the symmetry axis
($y=\pm1, x=\displaystyle{\frac zM}$) with (\ref{ques}), we conclude
that:
\begin{eqnarray}
& &h_{2n+1}^q=0 \quad \forall n\geq 0 \\ \nonumber & &h_0^q=0\\
\nonumber & &h_{2n}^q = -(4n+1) \sum_{j=1}^n I_j^q L_{2n,2j} \quad
\forall n\geq 1 \ .
\end{eqnarray}
Using (\ref{iq}), it can be verified that in the expression for
the constants $h_{2n}^q$, i.e.,
\begin{equation}
h_{2n}^q = -(4n+1) [\sum_{j=0}^n \frac 54 L_{2n,2j} -\sum_{j=0}^n\frac
{15}{8} \frac{L_{2n,2j}}{2j+1} + \sum_{j=0}^n \frac {15}{8}
\frac{L_{2n,2j}}{2j+3}] \ ,
\label{haches}
\end{equation}
for $n\geq 2$ the two last terms are equal to zero (see Lemma 2 of
the appendix in \cite{y2}), and taking into account the orthonormality of
Legendre's polynomials we have:
\begin{equation}
h_{2n}^q = -\frac 54 (4n+1) \quad , \quad \forall n\geq 2 \ ,
\end{equation}
whereas for $n=1$ we see that in (\ref{haches}) the last term is not
zero (by virtue of Lemma 2 in \cite{y2}) and therefore:
\begin{equation}
h_2^q = -\frac {15}{2} \ .
\end{equation}

Once we have determined the constants $h_{2n}^q$, we can state that the
contribution  on the parameter $q$ to the solution is given by the
potential
\begin{equation}
\zeta_2 = -\frac{15}{2} Q_2(x) P_2(y) - \sum_{n=2}^{\infty} \frac 54
(4n+1) Q_{2n}(x) P_{2n}(y) \ .
\end{equation}
Alternatively, by using the Heine identity \cite{hein}, \cite{forts}(see also equation (72)
in \cite{y2}) we have:
\begin{eqnarray}
\zeta_2 = \frac{5}{4} Q_0(x) P_0(y) -\frac{5}{4} Q_2(x) P_2(y)-
\sum_{n=0}^{\infty} \frac 54 (4n+1) Q_{2n}(x) P_{2n}(y)\\ \nonumber =
\frac{5}{4} Q_0(x) P_0(y) -\frac{5}{4} Q_2(x) P_2(y) -\frac 54
\frac{x}{x^2-y^2} \ .
\end{eqnarray}

\subsection{Analysis of the solution}

We shall make some comments on the approximate solution obtained \footnote{Note that for a real variable $x>1$, we have \cite{ince} the following associated Legendre functions of the second kind: $Q_0(x)=-\frac 12 \ln \frac{x-1}{x+1}$, $Q_2(x) = -\frac 12 P_2(x) \ln \frac{x-1}{x+1} -\frac 32 x$}:
\begin{equation}
\xi = \frac 1x+\frac{y}{x^2} {\cal J} +q \frac{1-x^2}{x^2} \frac 54 [
Q_0(x) P_0(y) - Q_2(x) P_2(y) -\frac{x}{x^2-y^2}]
\label{solution}
\end{equation}
As is well known, the general stationary axisymmetric gravitational field can be described by the line element \cite{forts}
\begin{eqnarray}
ds^2 & = & -f (dt -\omega d \phi)^2 + \\ \nonumber
& & \sigma^2 f^{-1} [e^{2 \gamma} (x^2-y^2) (\frac{dx^2}{x^2-1}+\frac{dy^2}{1-y^2})+(x^2-1)(1-y^2) d\phi^2 ] \ ,
\end{eqnarray}
where the coordinate system is $\{t,x,y,\phi\}$, the space-like coordinates $\{x,y\}$ being the prolate spheroidal coordinates, $x$ represents a radial coordinate, whereas the other coordinate $y$ represents the cosine function of the polar angle, with $\sigma$ a real constant that, for our case,  can be identified with the mass; the unknown functions $f$, $\gamma$ and $\omega$ only depend on $x$ and $y$.

The Ernst potential \cite{forts} $E=\frac{1-\xi}{1+\xi}$ is defined by $E=f+i W$, where 
\begin{eqnarray}
W_{,x} & = & \sigma^{-1} (x^2-1)^{-1} f^2 \omega_{,y} \\ \nonumber
W_{,y} & = & -\sigma^{-1} (1-y^2)^{-1}f^2 \omega_{,x} \ ,
\label{omg}
\end{eqnarray}
and the function $\gamma$ satisfies the following partial differential equations:
\begin{eqnarray}
&\gamma_{,x}  =  \displaystyle{\frac {1-y^2}{4 f^2 (x^2-y^2)}} \times \\ \nonumber
&  [ x(x^2-1) E_{,x} E_{,x}^{\star}-x(1-y^2) E_{,y} E_{,y}^{\star} - y(x^2-1) (E_{,x} E_{,y}^{\star}+  E_{,y} E_{,x}^{\star})] \\ \nonumber
& \gamma_{,y}  = \displaystyle{ \frac {x^2-1}{4 f^2 (x^2-y^2)}} \times \\ \nonumber
&  [ y(x^2-1) E_{,x} E_{,x}^{\star} -y(1-y^2)E_{,y} E_{,y}^{\star} +x(1-y^2)(E_{,x} E_{,y}^{\star}+  E_{,y} E_{,x}^{\star}) ] \ .
\label{gam}
\end{eqnarray}
Hence, from (\ref{solution}) we obtain:
\begin{eqnarray}
f & = & \frac{x-1}{x+1} \{1+\frac 52 q  [
Q_0(x) P_0(y) - Q_2(x) P_2(y) -\frac{x}{x^2-y^2}]\} \\ 
W & = & -\frac{2 y}{(1+x)^2} \frac{J}{M^2}
\end{eqnarray}
and, therefore, by solving equations (\ref{omg}) and (\ref{gam}) up to order $q$ and ${\cal J}$ we have:
\begin{eqnarray}
& &\omega  =\displaystyle{2 \frac JM \frac{y^2-1}{x-1}} \\
& &\gamma  = \displaystyle{ \frac 12 \ln(\frac{x^2-1}{x^2-y^2}) - \frac {15}{8} q  x (1-y^2) \ln(\frac{x-1}{x+1}) +} \\ 
& & \qquad \displaystyle{ \frac 58 q [6 y^2 -\frac{2}{(x^2-y^2)^2} (y^2+x^2(x^2-3 y^2 +1))-4] \ . \nonumber}
\end{eqnarray}
 
The solution fulfils asymptotical flatness conditions since all its metric functions have the good asymptotic behaviour \cite{forts},  and the metric does not possess any singularity on the symmetry axis, i.e., $\displaystyle{\lim_{x->\infty} f =1}$, $\displaystyle{\lim_{x->\infty} \omega = 0}$, $\displaystyle{\lim_{x->\infty} \gamma = 0}$,  $\displaystyle{\lim_{y^2->1} \gamma =0}$.

It is possible to  check the above-mentioned good quality of the solution for
describing the approximate gravitational field of a mass with only
quadrupole and angular momentum. First, if we calculate the multipole
moments of the solution we obtain:
\begin{eqnarray}
& &M_0= M \quad , \quad M_1={\cal J} M^2\equiv i J \quad , \quad M_2=q M^3\equiv Q \\ \nonumber&
& M_3=0 \quad , \quad M_4={\cal J}^2 \frac{M^5}{7} \\ \nonumber& &M_5=\frac{M^6}{21} ({\cal J} q 8 -
{\cal J}^3)\\ \nonumber& &M_6=\frac{M^7}{21} ({\cal J}^2 -\frac{51}{11}
q^2+\frac{23}{11} q {\cal J}^2)\ .
\end{eqnarray}

These expressions show that the solution possess equatorial symmetry since massive multipole moments of odd order are null whereas dynamic multipole moments of even order are equal to zero.

As can be seen, all multipole moments are several quantities of order higher
than ${\cal J}$, $q$, or products of both; i.e. the order of
approximation considered. 
Take into account that both ${\cal J}$ and $q$ are very small quantities, since they are the dimensionless parameters $i J/M^2$ and $Q/M^3$ respectively, and we are considering small deviations from sphericity, i.e., quadrupole moment $Q$ and angular momentum $J$ are less than the first multipolar order, the mass $M$.
Obviously, this result was expected because
the solution is constructed in such a way that the coefficients $m_n$ are
taken up to a linear contribution in both parameters.

Furthermore, this solution has good limits when we consider
the parameter $q$ or ${\cal J}$ to be equal to zero. Of course, $q={\cal
J}=0$ reproduces the Schwarzschild solution, as we already mentioned,
since the solution is constructed by starting from $\xi_0$, the
Ernst potential corresponding to spherical symmetry space-time. If
we take in (\ref{solution}) $q=0$, the resulting solution should
represent the first correction to the static configuration of the
Schwarzschild solution due to the effect of rotation. In fact, that
solution, (\ref{kerr1orden}), corresponds to the Kerr solution up to
order ${\cal J}$ in the expansion of the Ernst potential in power
series of that parameter \cite{y2}, i.e.,
\begin{equation}
\xi^{kerr}=\frac{1}{x \sqrt{1+{\cal J}^2}-{\cal J} y} \ ,
\end{equation}
whose expansion in power series of parameter ${\cal J}$ gives:
\begin{eqnarray}
\xi_0^{kerr}&=& \frac 1x \\ \nonumber 
\xi_1^{kerr}&=&
\frac{x^2-1}{x^2} Q_1^{(2)}(x)P_1(y)=\frac 12 \frac{y}{x^2} 
\end{eqnarray}

Finally, if we take ${\cal J}=0$ in (\ref{solution}), the
corresponding solution
\begin{equation}
\xi = \frac 1x+q \frac{1-x^2}{x^2} \frac 54 [ Q_0(x) P_0(y) - Q_2(x)
P_2(y) -\frac{x}{x^2-y^2}]
\label{soluq}
\end{equation}should represent the first contribution to the deformation
 from the spherical symmetry configuration due to the quadrupole moment.
 Therefore, it should be the $MQ^{(1)}$ solution  \cite{y0}, \cite{y1}
 constructed with this claim. In order to verify this conjecture, 
it is necessary to check that the metric function $\Psi$ corresponding
 to the $MQ^{(1)}$ solution   generates the Ernst potential (\ref{soluq}) up to
 order $q$. Since the $MQ^{(1)}$ solution is given by 
$\Psi=\Psi_0+q \Psi_1$, we have:
\begin{equation}
E\equiv e^{2 \Psi} = A (1+2 \Psi_1 q)+{\cal O}(q^2) \ ,
\end{equation}
with $A \equiv e^{2 \Psi_0}=\frac{x-1}{x+1}$, and,
\begin{equation}
\xi =\frac{1-E}{1+E} = \frac{B-D}{B+D} 	 ,
\end{equation}
with the following notation: $B \equiv 1-A$, $C\equiv 1+A$ and $D \equiv 2 A q \Psi_1$.
Hence, up to order $q$, we have that
\begin{equation}
\xi^{MQ^{(1)}} \approx \frac BC -\frac DC (\frac BC +1) = \frac 1x - \frac{x^2-1}{x^2} q \Psi_1+{\cal O}(q^2) \ .
\label{ximq1}
\end{equation}
By substituting the expression for $\Psi_1$ \cite{y1} it is clear that (\ref{ximq1}) is actually
equivalent to  the solution obtained (\ref{solution}) up to order $q$, i.e., (\ref{soluq}).

\section{Conclusion}

In previous sections an approximate solution of the vacuum Einstein
field equations with axial symmetry was obtained. The metric functions of
the Weyl line element are given explicitly by means of the
corresponding Ernst potential. To construct the solution we have made use of the
relationship between the coefficients of the Ernst potential appearing at
its series expansion in powers of the Weyl coordinate along the
symmetry axis, and the relativistic multipole moments of the solution.

The relevance of this new solution, in contrast to other known
solutions, has a very interesting physical meaning since it can be used
to describe -in the perturbative sense- the corrections to the spherical
symmetry that an angular and quadrupole moment would incorporate to the
solution.  As already seen, the solution has two independent
parameters, $q$ and ${\cal J}$, that are  directly related to these multipole
moments and, by using those parameters, the approximation up to first
order of the pure multipole $MJQ$ solution is constructed.  The good limits of the solution for the cases $q=0$ and ${\cal
J}=0$ are verified, thereby recovering  the  $MJ$ solution up to first order (which is
exactly Kerr at this order) and the  $MQ^{(1)}$ solution  respectively.

A significant difference of this solution with respect to the Kerr
solution  is that it allows us to control the magnitude of the
contributions from the quadrupole deformation and the effect of
rotation independently by means of its two parameters. The
different behaviour of test particles moving along geodesics in the
radial direction of the $MQ^{(1)}$  solution with respect to that
movement in the case of spherical collapse is already known \cite{colapso}. We 
expect that this solution may be a relevant reference to describe 
non-spherical collapse since it can supply more detailed information
about the process.  Study of all the implications of this claim for the solution
is still in progress and  will be addressed in  future work, where
the good behaviour of the event horizon will be shown as well as a
detailed study of the test particle geodesics. In any case, it should be stressed that the approximation used to construct the solution is
correct in the sense that it allows one to approach the source without any
type of discontinuity, and moreso for the strong field case (where the
approximation is better  since the parameters $q \equiv Q/M^3$
and ${\cal J} \equiv i J/M^2$ become sufficiently less than 1 to be
neglected).

Apart from achieving an interior solution for the $MQ^{(1)}$ \cite{luisbarreto}, 
it is expected that a search for an interior solution matching this new solution would provide a 
very interesting global model for the description of compact bodies slightly different from the spherical, but never realistic, source.

\vbox{}

\section*{Acknowledgments}
This work was partially supported by the Spanish Ministry of Science
and Technology under Research Proyect No. BFM 2003-02121, and the
Grant No SA002/03 from  Junta de Castilla y Le\'on.

I also wish to thank Professor Luis Herrera for his helpful discussions and
comments, as well as Dr. M.A. Gonz\'alez Le\'on and Dr. A. Alonso Izquierdo for their help and comments on the final version of the work.

\end{document}